\documentclass[twocolumn]{aastex6}
\usepackage{graphicx}	
\usepackage{amsmath}
\usepackage{amssymb}
\usepackage{subfigure}
\usepackage{color}

\usepackage{newtxmath}
\usepackage{amssymb, braket}
\usepackage{hyperref}
\usepackage{comment}
\usepackage{natbib}
\usepackage{ipa}

\newcommand\simlt{\lower.5ex\hbox{$\; \buildrel < \over \sim \;$}}
\newcommand\simgt{\lower.5ex\hbox{$\; \buildrel > \over \sim \;$}}

\begin{document}

\title{The Chiral Puzzle of Life}
\author{Noemie Globus\altaffilmark{1,2} and Roger D. Blandford\altaffilmark{3}}
\altaffiltext{1}{Center for Cosmology \& Particle Physics, New-York University, New-York, NY10003, USA. E-mail:  globus@nyu.edu}
\altaffiltext{2}{Center for Computational Astrophysics, Flatiron Institute, Simons Foundation, New-York, NY10003, USA}
\altaffiltext{3}{Kavli Institute for Particle Astrophysics \& Cosmology, Stanford University, Stanford, CA 94305, USA. E-mail:  rdb3@stanford.edu}

\begin{abstract}
Biological molecules chose one of two structurally, chiral systems  which are related by reflection in a mirror. It is proposed  that this choice was made, causally, by cosmic rays, which are known to have a large role in mutagenesis. It is shown that magnetically-polarized cosmic rays, that dominate at ground level today, can impose a small, but persistent, chiral bias in the rate at which they induce structural changes in simple,  chiral monomers that are the building blocks of biopolymers. A much larger effect should be present with helical biopolymers, in particular, those that may have been the progenitors of RNA and DNA. It is shown that the interaction can be both electrostatic, just involving the molecular electric field, and electromagnetic, also involving a magnetic field. It is argued that this bias can lead to the emergence of a single, chiral life form over an evolutionary timescale. If this mechanism dominates, then the handedness of living systems should be universal. Experiments are proposed to assess the efficacy of this process.
   
\end{abstract}

\section*{Introduction}

Living organisms comprise a system of molecules organized with specific handedness. Handedness - or chirality - is, following Kelvin's original definition, the geometric property of an object that cannot be superimposed on its mirror image \citep{kelvin1894}. 
In chemistry,  mirror images of the same chiral molecule are called enantiomers\footnote{from the Greek $\epsilon\chi\theta\rho$\'o$\varsigma$, "enemy" or "opposite".}. Both share
the same chemical characteristics.

\begin{figure*}
\centering
\includegraphics[scale=0.7]{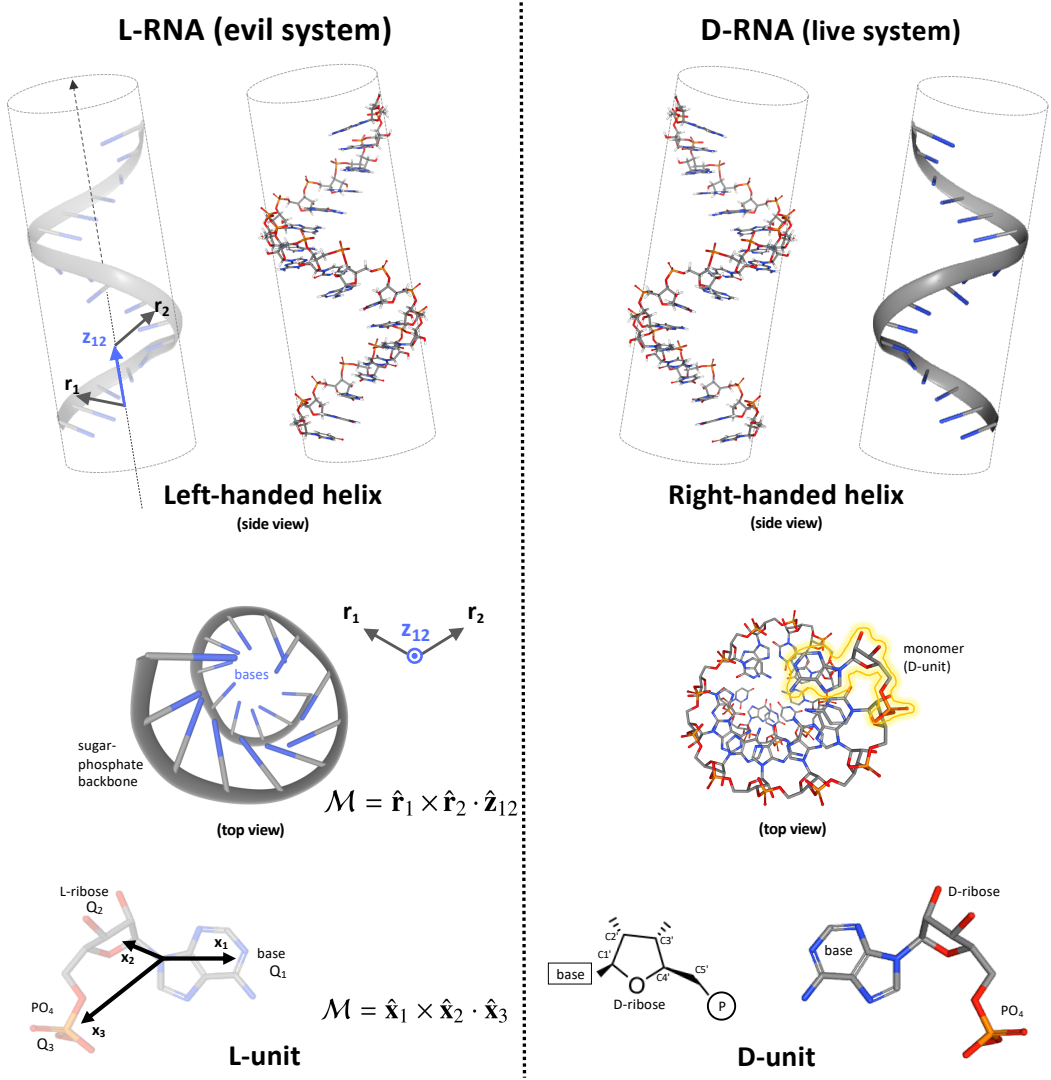}
\caption{ \footnotesize The figure shows the 3D structure of the RNA molecule and its mirror image.   The direction of the helical conformation of the nucleic acids derives from the underlying chemical chirality of the sugar backbone. The nucleic acids  contain only  right-handed sugars (D-ribose in RNA, D-deoxyribose in DNA), shown in the right-hand side of the figure. They naturally assume a right-handed helical conformation.  
In the mirror world (left-hand side in the figure), the nucleic acids would contain only left-handed sugars (L-ribose or L-deoxyribose) and would assume a predominantly  left-handed helical conformation.
}
\label{fig:live_evil}
\end{figure*}

The ribonucleic and deoxyribonucleic acids (RNA and DNA), responsible for the replication and storage of genetic information, are made up of linear sequences of  building blocks with the same handedness, called nucleotides, whose arrangement is neither periodic nor random and contains the genetic information needed to sustain life \citep{schrodinger1944,shannon1948, watson1953, shinitzky2007}. The chirality of the nucleotides confers helical structure on nucleic acids. Nucleic acids are very large molecules and the torsional angles between the chiral units vary systematically, as exhibited by the Ramachandran plot \citep{keating2011} which demonstrates that even a flexible biopolymer retains chirality.  As RNA and DNA are made of D-sugars (right-handed, by human convention), the more stable conformation is a right-handed helix (see Figure~\ref{fig:live_evil}). The homochirality of the sugars has important consequences for the stability of the helix, and hence, on the fidelity or error control of the genetic code. All the twenty encoded amino acids are left-handed (again by human convention). Sometimes, both enantiomers of the same molecule are used by living organisms, but not in the same quantity and they perform different tasks. 

While DNA/RNA-based life, as observed so far, has clearly chosen one functional chirality, which we call ``live'', the alternative choice, which we call ``evil'', could have developed along a separate, synchronized path making similar evolutionary choices in response to changes in common environments, except for very small effects which are the main topic of this Letter. 
However, a precise equilibrium between the two choices seems quite unlikely given the high replication rate. There is a small entropic price, but this is surely paid by the greater facility of storing information and the higher reliability of the replication \citep{schrodinger1944}. 

For DNA today, radiation increases the frequency of gene mutations; this has been known since the pioneering work of \citep{muller1927} that showed that the mutation rate is proportional to the radiation dose, much of it attributable to ionization by cosmic rays. 
The muon component dominates the flux of particles on the ground at energies above 100 MeV, contributing 85\% of the radiation dose from cosmic rays \citep{atri2011}. Muons have an energy sufficient to penetrate considerable depths, and they are, on average, spin-polarized. Ionization by spin-polarized radiation could be  enantioselective \citep{zeldovich1977}. Therefore, we argue that the mutation rate of live and evil organisms would be different. As there could be billions or even trillions of generation of the earliest and simplest life forms, a small difference in the mutation rate could easily sustain one of the two early, chiral choices.

When Pasteur discovered biological homochirality, he recognized it as a consequence of some asymmetry in the laws of nature: {\it "If the foundations of life are dissymmetric, then because of dissymmetric cosmic forces operating at their origin; this, I think, is one of the links between the life on this earth and the cosmos, that is the totality of forces in the universe"} \citep{pasteur1848, quack1989}. Had Pasteur been alive a century later, the discovery of parity violation in the weak interaction \citep{lee1956,wu1957} would have strengthened his view. An object exhibits {\it physical} chirality when its mirror image does not exist in Nature, as a consequence of parity violation in the weak interaction. The result of applying the parity operation on an elementary weak process, such as the decay, $\pi^+\rightarrow\mu^++\nu_\mu$, is not found in Nature because neutrinos are chiral particles. 
In the language of quantum mechanics, the basic Hamiltonian of a chiral molecule does not commute with the parity operator and, if we include weak neutral currents, there will be a parity-violating energy difference (PVED) between the two enantiomers \citep{yamagata1966}. However it is extremely small, $\sim10^{-17}kT$ in water \citep{salam1991} and larger consequences of chirality must be sought.
While the effectivity of PVED in generating biological homochirality is still under debate, some authors have attempted to work with this small PVED and showed that it may theoretically suffice to bring strong chiral selectivity \citep{kondepudi1984}. An enantiomeric excess due to neutral weak currents has been reported in crystalline materials \citep{szabo1999}.

In a beautiful paper, Pierre Curie  addressed the question of  chirality transfer from light to molecules, specifically involving circular polarisation \citep{curie1894}.  The sense of the rotation reflects the underlying chirality of the molecules, though the relationship is not simple and depends upon the wavelength of the light (Optical Rotatory Dispersion). This rotation can be accompanied by a difference in the absorption (Circular Dichroism), consistent with the Kramers-Kronig relations \citep{kramers1927,kronig1926}. On this basis, it has been suggested that a specific source of circularly polarized light (CPL) might favor one set of enantiomers over the other \citep{bailey1998}.

Laboratory experiments have demonstrated that it is possible to induce an enantiomeric excess of amino acids by irradiation of interstellar ice analogs with UV CPL \citep{demarcellus2011}. However, this raises two problems. Firstly, Circular Dichroism is also wavelength-, pH- and molecule-specific \citep{hendecourt2019}. It is hard to see how one sense of circular polarization can enforce a consistent chiral bias, given the large range of environments in which the molecules are found. Secondly, it is often supposed that astronomical sources supply the polarization. However, optical polarimetry within the Galaxy reveals no consistent sense of circular polarisation and the observed degrees of polarization in the UV are generally quite small \citep{bailey2001}. 

If we seek a  universal, chiral light source, that consistently emits one polarization over another, then we are again drawn to the weak interaction in order to account for a universal asymmetry. One option is to invoke spin-polarized particles, which can radiate one sense of circular polarization through {\v C}erenkov radiation or bremsstrahlung and can preferentially photolyze chiral molecules of one handedness \citep{vester1959,lahoti1977,gusev2019}. Another option is to invoke supernova neutrinos \citep{boyd2018}. However, the small chiral bias is unlikely to lead to a homochiral state and some pre-biotic amplification mechanism is still required. This suggests considering, instead, enantioselective bias in the evolution of the two living systems. 

\section*{Molecular chirality of biomolecules}
Consider, first, a model of a small chiral molecule, part of a larger helical polymer (see Fig.\ref{fig:live_evil}), which we idealize as an unequal tripod. There is a vertex or ``target'' at the origin and three distinguishable atoms or groups with position vectors ${\bf x}_1$, ${\bf x}_2$, ${\bf x}_3$. There is a classical electrostatic field associated with the point charges at these four sites. It is helpful to introduce a pseudoscalar (changes sign under reflection)  ``molecular chirality'', ${\cal M}$, which has to change sign upon reflection and a clear choice is ${\cal M}_{\rm tripod}=\hat{\bf x}_1\times\hat{\bf x}_2\cdot\hat{\bf x}_3$.

A second simple, semi-classical model has electrical charge and current confined to the surface of a sphere surrounding a central, nuclear charge which is canceled by the net charge on the sphere. The current allows for an electromagnetic chirality, with the simplest expression ${\cal M}_{\rm em}=\hat{\bf d}\cdot\hat{\bf m}$.These two models are appropriate for small molecules or monomers that are the constituents of naturally helical biopolymers. 

Our third, simple model involves a cylindrical, electrostatic potential, like  a ``Barber pole'', \citep[cf.][]{wagner1997}, $\Phi=R_0(r)+R_1(r)\cos(kz-m\phi)$ with $k>0$. In this case, the molecular chirality ${\cal M}$ can be chosen as ${\cal M}_{\rm helix}=m=+1\,(-1)$ for a live (evil) molecule. These definitions of ${\cal M}$ are illustrated in Fig.~\ref{fig:live_evil}.
\newpage
\section*{Cosmic-ray lodacity}
\begin{figure*}
\centering
\includegraphics[scale=0.4]{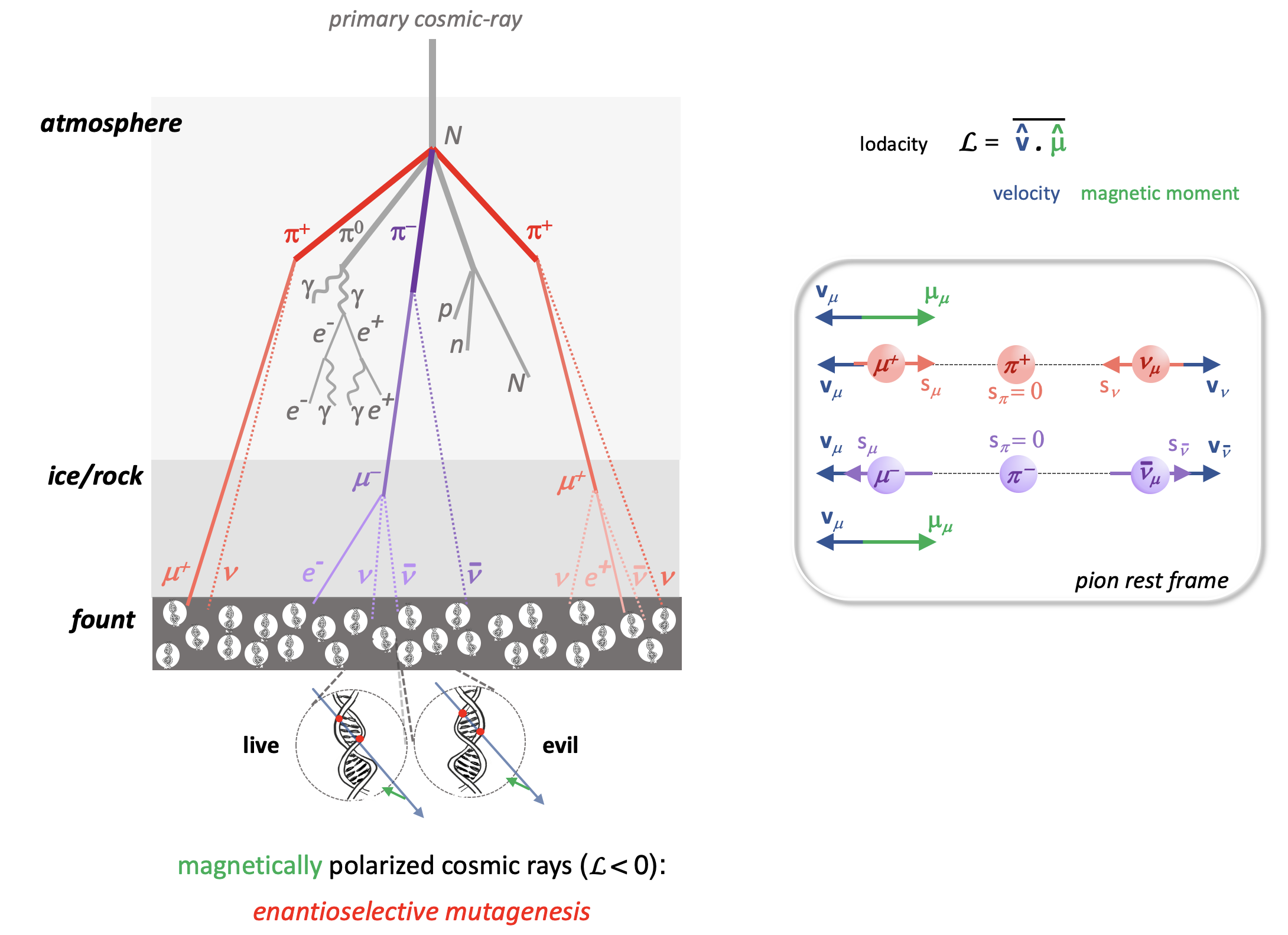}
\caption{
\footnotesize 
The chiral quantity associated with cosmic-ray showers is the lodacity ${\cal L}=\overline{\hat{\boldsymbol{\muup}}\cdot\hat{\textbf v}}$.
Spin-polarized muons (respectively antimuons) and their daughter electrons (respectively positrons)  are produced in air showers mainly from charged pion decay. They are indicated in color in the sketch (including the parent charged pions) while the non spin-polarized (electromagnetic and nucleonic) components are in grey. The muon component dominates at ground level but will slow and decay into electrons and positrons below ground level. The charge parity (CP) invariance leads to an universal sign of the cosmic-ray lodacity ${\cal L}<0$. 
}
\label{fig:demeter}
\end{figure*}

Charged, cosmic ray protons, with energies just above the threshold for pion production, collide with nitrogen and oxygen nuclei in the upper atmosphere to create $\pi^+$, $\pi^-$  \citep{gaisser2012}. The $\pi^+$ decay within a few meters into $\mu^+$ with half life $\sim2\,\mu{\rm s}$, which decay, in turn into $e^+$. As pions are spinless and the decays are weak, the $\mu^+$ and $e^+$, spin directions $\hat{\bf s}_{\mu,e}$, are preferentially anti-aligned with their direction of motion, $\hat{\bf v}$ in order to balance the antiparallel spins of the accompanying neutrinos (Fig.~\ref{fig:demeter}). The associated magnetic dipole moments are given by $\boldsymbol{\muup}_{\mu,e}=e\hbar\hat{\bf s}_{\mu,e}/2m_{\mu,e}$. The $\pi^-$ decay into $\mu^-$, $e^-$ with preferentially aligned spins but with magnetic moments also anti-aligned with $\hat{\bf v}$ (see Fig.~\ref{fig:demeter} and appendix \ref{ASA}).
 
 We introduce a pseudoscalar quantity, ``lodacity'' (after lodestone) to express the physical chirality of the cosmic rays. This is defined by 
\begin{equation}\label{eq:lodacity}
{\cal L}_i(T)=\overline{\hat{\boldsymbol{\muup}}\cdot\hat{\textbf v}},
\end{equation}
where we average over all cosmic rays of type $i$ and kinetic energy $T$.  Well above threshold, ${\cal L}\sim-1$ for freshly created $\mu$, and ${\cal L}\sim-0.3$ for new $e$, in the pion rest frame (see supplementary materials).  ${\cal L}$ will be further degraded as the cosmic rays lose energy through scattering electrons with ${\cal L}\propto v$ roughly. In addition the secondary electrons will be mostly unpolarized and further diminish the lodacity of the cosmic rays that irradiate the molecules. 

At sea-level today, most cosmic rays are muons with an average flux $\sim160\,{\rm m}^{-2}\,{\rm s}^{-1}$  \citep{lipari1993}. However, the flux and the atmosphere could have been quite different; the young sun and its wind are likely to have been much more active.
The protobiological site, which we call the ``fount'', may have been below rock, water or ice which can change the shower properties and lodacity.

\section*{Enantioselective interaction}
We now turn to the interaction that couples the cosmic ray lodacity ${\cal L}$ to the molecular chirality ${\cal M}$. We seek an effect that is proportional to the product ${\cal LM}$ which will distinguish live and evil molecules exposed to the same cosmic ray flux and will be unchanged upon reflection - a chiral bias. This effect must be translated into a difference in the ultimate mutation rate, a pathway that is poorly understood even in contemporary biology. A high energy particle can excite an electron locally \citep{rosenfeld1928}. Typically, the de-excitation is fast and radiationless and involves vibrational and rotational modes. This ``internal conversion'' can therefore cause local structural change in the molecule. Cosmic radiation also induce ionization which introduces changes in the electronic structure of the biomolecules and can lead to mutations. DNA, today, is presumably a far less error-prone copier than the first genetic biopolymer. Repair may also have been a factor when life began.

The cosmic rays themselves are supposed here to be spatially homogeneous and isotropically distributed with respect to the molecules. Their cosmic ray-averaged magnetic moment is also presumed to be strictly antiparallel to $\bf v$ although scattering processes, or an external magnetic field, can introduce an angle between $\bf v$ and $\boldsymbol{\muup}$. This is important because the chiral part of the electrostatic interaction involves a force given by ${\bf v}\cdot\boldsymbol{\muup}\times\nabla{\bf E}$ which vanishes unless the velocity is perturbed. Details of the calculation of the chiral bias $\delta \ln P$ are presented in appendix \ref{sec-chiralpropa}.

We start with the unequal tripod (as illustrated by Fig. \ref{monomer_electric} in the Supplements). Consider a cosmic ray with charge $qe$ mass $Mm_e$, subrelativistic velocity $\bf v$ and impact parameter vector with respect to the target at the origin given by $\bf b$. The trajectory will be linearly perturbed by the Coulomb force due to the the charge $Q_1e$ at ${\bf x}_1$. This will cause a velocity perturbation $\delta{\bf v}_1$, which creates a second order chiral force in combination with the electric field from the second atom. This produces a displacement at the target and the gradient of the displacement is equivalent to a chiral change in the particle flux. However, this change vanishes after we average over $\hat{\bf v}$. This is expected because we have only involved two of the atoms. We have to go to third order perturbations to get an average chiral difference. Furthermore, the chiral bias vanishes if two of the tripod legs are of equal length. This is also be expected because the charges $Q_i$ are multiplicative and if the bonds are of equal length then the geometrical structure by itself is not chiral. If the probability of a mutation is $P$ and the difference between this probability for live and evil molecules is $\delta P$, then $\delta\ln P\sim\alpha^7{\cal LM}_{\rm tripod}q^2Q^3M^{-4}(c/v)^5$  (derived in the Supplements), where $\alpha\sim0.0073$ is the fine structure constant. This is too small to be of interest but does bring out clearly the factors that are important in a larger effect. 

Next, consider the sphere with surface charge and current (as illustrated by Fig. \ref{monomer_em} in the Supplements). The simplest and largest chiral effect is electromagnetic and comes from combining its electric and magnetic dipole moments.  In this case $\delta\ln P\sim\alpha^4{\cal LM_{\rm em}}qM^{-2}(c/v)^2$. A similar conclusion was reached through a quite different argument, by  \cite{zeldovich1977}. However, there seems to be no good reason why ${\cal M}_{\rm em}$ should be non-zero. 

The third, electrostatic helical model is also chiral (as illustrated by Fig. \ref{barberpole} in the Supplements).
We invoke an isotropic ``mutability'', $\kappa$, which is the probability per unit length of cosmic ray trajectory through the molecule that a significant mutation will result. We suppose that the mutability $\kappa$ has a both a radial and a helical component, like the electrostatic potential. We find that the third order chiral bias comprises a sum of terms that contain two helical factors and one radial factor. If the structure is at all similar to RNA then it is likely that the bias is dominated by terms with an axisymmetric mutability, $\kappa$. We find that the original cosmic ray positrons, which outnumber the electrons, are deflected radially inward when encountering a live molecule and outward with an evil molecule. This implies that  interactions with the nucleobases must cause more mutations than those with the suger-phosphate backbone. The overall chiral bias is given by $\delta\ln P\sim \alpha^5{\cal LM}qM^{-3}(c/v)^3$.

Finally, we consider an electromagnetic helical model where we suppose that if the individual monomers carry magnetic dipoles as well as electric dipoles. Then, although the magnets do not line up as in a ferromagnet, there may be enough near neighbor correlation for there to be an electromagnetic, chiral bias which could be $\sim(v/\alpha c)$ times the electrostatic bias.

The bias $\delta \ln P$ ($\sim10^{-7}$ for keV electrons, times the lodacity and the fractional difference between positive and negative charges in the barber pole model) is the relative difference in the mutation rate between live and evil organisms. In appendix \ref{sec-evolampli}, we use the the logistic equations to model the population growth. Starting with a racemic mix the enantiomeric  excess is  $e.e.={\rm tanh}(\delta \ln P\, T/2)$ where $T$ is the time multiplied by the growth rate. If we add a balanced, live-evil ``conflict'' then homochiralization is speeded up. Either way, a small bias in the mutation rate can achieve this on an evolutionary timescale.

\section*{Discussion}

In this paper, we have proposed that homochirality is a deterministic consequence of the weak interaction, expressed by cosmic irradiation of helical biopolymers which may have affected the way they fold or assemble to make the first living organisms. This is the consequence of the coupling $\cal M L$ that can lead to  symmetry-breaking as anticipated by Pasteur. The choice that was made is then traceable to the preponderance of baryons over antibaryons, established in the early universe and ultimately to the symmetries of fundamental particle interactions presenting requirements (including leptonic CP violation)  as first elucidated by  \citep{sakharov1967}. We have also demonstrated how this quite small chiral bias can lead to homochirality after sufficient generations of self-replicating molecules and shown how conflict can speed up this Manichean struggle. 

Much more study is needed to determine if these processes suffice to account for homochirality. In particular, it will be necessary to investigate different shower models to understand the evolution of the lodacity and to develop a quantum mechanical model of collisional excitation. There are many other effects to explore. For example, it has been shown that the adsorption of chiral molecules on specific surfaces can   enhance the optical activity by several orders of magnitude because of the electric dipole - electric quadrupole interaction \citep{wu2017}. The interaction investigated may have analogs in biological environments.

A key issue facing astrobiology is assessing what subset of environments are necessary for the emergence of life. Good candidate environments for the polymerization of meteorite-delivered nucleobases are small, warm ponds produced by hydrothermal conditions associated with volcanic activity on early Earth \citep{pearce2017}.  Their wet and dry cycles have been shown to promote the polymerization of nucleotides into long chains \citep{da2015}. Any rocky planet, with active geological processes and water, has the potential to create life, because it is likely to support considerable environmental diversity; in particular, surface-based locales, including beaches and sea-ice interfaces \citep{stueken2013}.
\citet{lingam2018} outlined the biological consequences of tides in producing wet-dry cycles or providing biological rhythms in environments where the light-dark cycle is absent.
Irradiation by polarized radiation can only lead to small enantiomeric excess, and cannot explain the large excesses (15\%) found in meteorites and amplification mechanisms must be sought \citep{glavin2019}. Should the amino acids found on meteorites be biogenic, they would have existed long before the appearance of life on Earth. 

Future space missions will return to Earth with samples collected on asteroids and on the martian sub-surface \citep{lauretta2017,yamaguchi2018,vago2017}. This will provide insight on the nature of the organic molecules and their chirality. As cosmic rays provide a natural connection between the weak interaction and living systems, we predict that, if ever indigenous biopolymers are found ({\textit i.e.} traces of living systems), they will have the same handedness as life on Earth. (Similar remarks apply to future samples returned from deep subterranean sites.)  Where life appeared first and  if cosmic rays played a role in its formation, are still open questions; but the important evolutionary consequences of spin-polarized cosmic radiation, which we propose here, is testable experimentally.

A prediction of our model is that the mutation rate is dependent upon the spin-polarization of the radiation. A possible experiment would be to measure the mutation rate of two cultures of bacteria under spin-polarized radiation (either  $e^\pm$ or $\mu^\pm$) of different lodacity with energy above the threshold necessary to induce double strand breaks in DNA ($\sim50$ eV). If the coupling between lodacity and molecular chirality is efficient in introducing a chiral bias, one of the two cultures should exhibit a much lower mutation rate. We emphasize that much  can be learned experimentally from the comparison of chiral molecules involved in biology and using both signs of lodacity which can be created at accelerators. It is not necessary to create ``mirror life'' to proceed. Once the dominant processes are identified, we can have confidence in our understanding of particle physics and quantum chemistry to draw the necessary conclusions. Additional experiments, relevant to our electromagnetic models, involve measuring the magnetic structure and properties of biopolymers. 

If these experiments show that the evolution of  bacteria is sensitive to polarization, this will be a good indication that magnetically-polarized cosmic rays are an important piece of the chiral puzzle of life. 

\section*{Acknowledgements}
The research of NG is supported by the Koret Foundation, New York University and the Simons Foundation. NG thanks Louis d'Hendecourt for discussions that inspired this work. The hospitality of the astrophysicists at RIKEN (ABBL, iTHEMS, r-EMU) during part of this work, is gratefully acknowledged.
We thank David Avnir, David Deamer, David Eichler, Anatoli Fedynitch, Peter Graham, Andrei Gruzinov, Ralph Pudritz, Stuart Reynolds, Jack Szostak and Jonas Lundeby Willadsen for helpful and instructive  comments.

\bibliography{scibib}

\onecolumngrid 
\appendix
\section{Air shower asymmetries}\label{ASA}
\subsection{Charge ratio}\label{charge_ratio}
Primary cosmic rays comprise mostly positive nucleons. This excess is transmitted via nuclear interactions to pions and then, on to muons.  The muon charge ratio is ${\cal R}_\mu\sim1.25$ below 1 TeV and increases to above $\sim1.4$ at higher energies \citep{gaisser2012}.  
Due to parity violation in the weak interaction, $\mu^\pm$ produced from decaying pions and kaons are on average spin-polarized. (The dominant contribution is from pion decay.) Their daughter electrons and positrons are also, on average, spin-polarized. The spin-polarized cosmic-rays can also produce UV CPL when propagating in the medium through emitting {\v C}erenkov radiation and bremsstrahlung.  

\subsection{Spin-polarized secondary particles}\label{spin-polarizedCR}
 
The spinless charged pion with a lifetime of 26 ns decays at rest into a left-handed muon neutrino and a muon:  $\pi^-\rightarrow \mu\bar{\nu}_\mu$ (and $\pi^+\rightarrow \mu^+\nu_\mu$ respectively). 
The pion has a mass of $m_\pi$ = 140 MeV/c$^2$, the muon has a mass $m_\mu$ = 106 MeV/c$^2$ and the neutrino is effectively massless.
We define $r_\pi=(m_\mu/m_\pi)^2$.
In the pion rest frame (denited by $*$), the momentum of the muon is 
\begin{equation}|\textbf p^*_{\mu}|=|\textbf p^*_{\nu}|=\frac{m_\pi c}{2}(1-r_\pi)\sim 29.8\, {\rm MeV/c} \end{equation}
and $E^*_\mu=\sqrt{{p^*_\mu}^2 c^2+m_\mu^2c^4}\sim109.8$ MeV.
Let the pion move in the laboratory with velocity $\textbf v_\pi/c={\beta}_\pi \textbf{e}_z$. 
Defining $\theta^*$, the angle of emission of the muon in the pion rest frame, we have the following relations for the muon momentum, energy, helicity ($h = \hat{\textbf{s}}\cdot\textbf{p}/|\textbf{p}|$) and angle of emission in the lab rest frame \citep{lipari1993}:
\begin{eqnarray} 
p_\mu=\gamma_\pi{p^*_\mu}\cos\theta^*+\beta_\pi\gamma_\pi{E^*_\mu}\,,\\
E_\mu=\gamma_\pi{E^*_\mu}+\beta_\pi\gamma_\pi{p^*_\mu}\cos\theta^*\,,\\
h(\beta_\pi,\theta^*)=\frac{1}{\beta_\mu}\left[ \frac{1-r_\pi+(1+r_\pi)\cos\theta^*\beta_\pi}{1+r_\pi+(1-r_\pi)\cos\theta^*\beta_\pi}\right]\,,\\
\tan\theta=\frac{\beta^*_\mu\sin\theta^*}{\gamma_\pi(\beta_\pi+\beta^*_\mu\cos\theta^*)}\,.
\end{eqnarray}
In the limit $\beta_\pi = 0$, the velocity of the muon is: $\beta_\mu = (1 - r_\pi)/ (1+r_\pi)\equiv\beta^*_\mu\sim0.27$  and we have $h$ = + 1 independent of the angle of emission of the muon. The polarization of the positive muon flux at sea level varies between $\sim30$\% and $\sim60$\%, depending on the energy, and is higher than the polarization of the negative muon flux \citep{lipari1993}. The lifetime of negative muons in matter is different because the negative muons interact with the nuclei of atoms, which will increase the charge ratio at greater depth.
In the same fashion, the electrons (positrons) from muon decay (antimuon decay) are mostly left-handed (right-handed) with the direction of the spin-aligned (opposite) to their momentum: $\mu^-\rightarrow e^-\nu_\mu\bar{\nu}_e$ ($\mu^+\rightarrow e^+\nu_e\bar{\nu}_\mu$).
The decay probability of a positron is $W(\theta)=(1+a\cos\theta)/(4\pi\tau_\mu)$ where $\theta$ is the angle between the spin direction and the positron trajectory, $\tau_\mu \sim 2.197 \,\mu$s is the mean lifetime, and the asymmetry term $a$ is a direct consequence that the muon decay is governed by the weak interaction, and depends on the positron energy, so the positron  angular distribution is ${\rm d}\Gamma/{\rm d}\cos\theta= W(\theta)$. The maximum and mean positron energies resulting from the three body decay are given by: $E_{{e^+}_{\rm max}}=(m_\mu^2+m_e^2)c^2/(2m_\mu^2)= 52.82$ MeV and $\bar{E}_{e^+}=36.9$ MeV. For a positron emitted with  energy of the order of $E_{{e^+}_{\rm max}}$, we have the maximum asymmetry $a=1$.   When averaged over all positron energies, $a=1/3$. 
 
\subsection{Circularly polarized radiation}
{\v C}erenkov radiation has a small degree of circular polarization which is dependent on the orientation of the spin of the initial particle. This is a purely relativistic quantum effect \citep{sokolov1940}. In the following we consider the difference between the number of left-handed photons and right-handed photons emitted from an electron of helicity $1/2$, {\it i.e.} spin  along its direction of motion, denoted by "$_1$".

Defining the ratio of the photon to the electron energies, $\xi= {\hbar\omega}/(2E_e)$, the velocity and Lorentz factor of the electron $\beta=v/c$, $\gamma=(1-\beta^2)^{-1/2}$, the {\v C}erenkov angle $\cos\theta_c=[1+\xi(n^2-1)]/(n\beta)$, and the function ${\cal F}=\cos\chi(\cos\theta_c-n\beta)+\gamma^{-1}\sin\chi\cos\phi\sin\theta_c$ where the angles are $\alpha=\pi/4$, $\chi=0$, $\phi=0$ for circular polarization, the number of right-handed photons $N_{1,+}$ (respectively left-handed $N_{1,-}$) is \citep{lahoti1977}
\begin{align}
N_{1,\pm}\propto 0.5(\beta\sin\theta_c)^2+\xi^2(n^2-1)\pm0.5(\beta\sin\theta_c)^2\cos(2\alpha)\mp\sin(2\alpha)\xi {\cal F}\,.
\end{align}
As an example, the ratio $(N_{1,+}-N_{1,-})/(N_{1,+}+N_{1,-})$  emitted by an electron  of energy $\sim0.8$ MeV ($\beta=0.77$), propagating in ice, is $\sim 1.3\,10^{-5}$ at a wavelength of 206 nm. (If the electron has helicity -1/2, $(N_{1,+}-N_{1,-})/(N_{1,+}+N_{1,-})$ has the same magnitude, but opposite sign). For muons at the same velocity ($\sim166$ MeV), the ratio is $1.8\,10^{-8}$ at the same wavelength. 

Longitudinally polarized $\beta$-radiation gives rise to circularly polarized bremsstrahlung. Using the Born approximation, \citet{mcVoy1957} derived the following formula for circular polarization in the limit where $E_e\sim h\nu$ and the emission angle of the photon $\theta=0^\circ$:
\begin{equation}
\frac{P_{\gamma}}{P_e}=\left(1+\frac{(1-\beta)(E_e+2mc^2)}{(2-\beta)E_e}\right)^{-1}\,.    
\end{equation}
Here $P_e$ is the polarization of the electron. 
The polarization transfer drops rapidly at electron energies $E_e$ below $\sim$1 MeV. Although the degree of circular polarization is quite small, $\sim 10^{-5}$ in the example above, it could still be large enough to impose homochirality if there is a general reason to couple this physical chirality to the geometrical chirality of biological molecules.  This matter deserves further consideration and, perhaps, experimental investigation.

\subsection{Lodacity Evolution}
\label{sssec:ESP}
As discussed previously, cosmic rays are preferentially positively-charged and they create $\mu^+$ and $e^+$ with lodacity ${\cal L}_i=\overline{\hat{\boldsymbol{\muup}}\cdot\hat{\textbf v}}<1$ for each species. This asymmetry can be degraded by three effects. The first is precession about an external magnetic field, $\textbf B_{\rm ext}$\footnote{Precession within the molecule is ignorable.}; the second is deflection of the particle momentum in a Coulomb interaction while leaving the direction of the magnetic moment unchanged.  This also leads to energy loss. The third is the dilution of the lodacity by unpolarized, ``knock on'' electrons. These are created during ionization loss when polarized cosmic rays collide with electrons in atoms. We consider these effects, in turn, for antimuons/muons and for positrons/electrons.

The Larmor radii of muons exceed their decay lengths so long as $B_{\rm ext}\lesssim1\,{\rm mT}$.  By contrast, nonrelativistic positron and electron Larmor radii $\sim(p/(m_{\rm e}c))(B_{\rm ext}/1\,{\rm mT})^{-1}\,{\rm m}$, where \textbf p is the electron momentum. This is quite likely to be small compared with their ranges and so the positrons and electrons will be channelled by an ordered magnetic field.

Under these circumstances, the equation of motion for a positron is 
\begin{equation}
\frac{d\textbf p}{dt}=\frac e{\gamma m_{\rm e}}\textbf{p}\times\textbf B_{\rm ext},
\end{equation}
where $\gamma=(1-p^2/m_{\rm e}^2c^2)^{-1/2}$. The magnetic moment will also precess about the magnetic field according to
\begin{equation}
\frac{d\boldsymbol{\muup}}{dt}=\frac e{\gamma m_{\rm e}}\boldsymbol{\muup}\times\textbf B_{\rm ext}.
\end{equation}

In the non-relativistic limit, which concerns us most, these equations then imply that \textbf p and $\boldsymbol{\muup}$ precess about $\textbf B_{\rm ext}$ with a common angular velocity $-eB_{\rm ext}/m_e$. We expect the spin-polarized daughter positrons to outnumber spin-polarized electrons of similar momenta and to be created with a momentum distribution that is axisymmetric about a downward direction, $\hat{\textbf g}\equiv g\hat{\textbf e}_z$. Furthermore, for each \textbf p, the distribution of $\boldsymbol{\muup}$ will be axisymmetric about $\textbf B_{\rm ext}$. For a given magnetic field direction, this can lead to an average spin/magnetic moment polarization projected perpendicular to the velocity. However, in this case, it is only $\cal L$ that has the required pseudoscalar form and the perpendicular component leads to no bias after full averaging. The precession contributes modest degradation of the mean polarization. 

Now turn to the cumulative effect of the deflections during Coulomb interactions. These are dominated by distant encounters and therefore are mostly small \citep[e.g.][]{2017mcp..book.....T}. A cosmic ray with momentum {\bf p} exchanges transverse momentum $\Delta{\bf p}_\perp$ with an individual electron effectively at rest, will lose kinetic energy $\Delta T=\Delta p_\perp^2/2m_e$, and be deflected through an angle $\Delta{\boldsymbol\theta}=\Delta{\bf p}_\perp/p$. Averaging over all encounters, we obtain
\begin{equation}
\left<\frac{d(\Delta\theta^2)}{d\Delta T}\right>=-\frac{2m_e}{p^2},
\end{equation}
for the sum of the mean square deflections along along two axes perpendicular to the momentum. It is helpful to introduce a quantity $\tau=\ln(1+2m_ec^2/T)$ which satisfies $d\tau/dT=-2m_e/p^2$. The mean square deflection angles add stochastically and so we can approximate individual deflections as small
and write
\begin{equation}
\left<\frac{d(\Delta\theta)^2}{d\tau}\right>=1.
\end{equation}
In addition, $<d\Delta\theta/d\tau>=0$.

We are now in a position to consider the evolution of a probability distribution function $P({\boldsymbol\theta},\tau)$ relative to the initial direction and the initial mean magnetic moment. It will satisfy a Fokker-Planck equation  \citep[e.g.][]{2017mcp..book.....T} with $\tau$ replacing time. 
\begin{equation}
\frac{\partial P}{\partial\tau}=\frac1{4\sin\theta}\frac\partial{\partial\theta}\sin\theta\frac{\partial P}{\partial\theta}.
\end{equation}
Multiplying this equation by $\cos\theta$ and integrating over solid angle, we obtain 
\begin{equation}
\frac{d<\cos\theta>}{d\tau}=\frac18\int_0^\pi d\theta\cos\theta\frac\partial{\partial\theta}\sin\theta\frac{\partial P}{\partial\theta}
\end{equation}
After integrating twice by parts, we obtain
\begin{equation}
\frac{d<\cos\theta>}{d\tau}=-\frac12<\cos\theta>,
\end{equation}
so that $<\muup_z>\propto\tau^{-1/2}$.

Now consider the evolution of the mean spin polarization of all secondary cosmic-rays of same mass $m_e$. If the cosmic-rays are created relativistically with lodacity ${\cal L}_0$, then, when they are nonrelativistic,
\begin{equation}\label{eq:LodT}
{\cal L}(T)\sim\,{\cal L}_0\left(\frac T{2m_ec^2}\right)^{1/2}.
\end{equation}

The third effect --- lodacity dilution by energetic knock on electrons --- is quite sensitive to the fount, the material that lies above it and the energies of particles that are most effective in bringing about mutation. The only particles that are of interest are those that are created close to the fount. An interesting complication for the electrons is that, as they are identical to particles with which they are colliding their total wave functions should be antisymmetric. This can introduce a spin-dependent interference term into the collision cross section \citep{messiah1981}. This effect can also be a factor in the quantum mechanical treatment of the direct interaction of an electron with the molecule.

A proper understanding of lodacity dilution requires shower simulations and a quantum chemistry treatment. These are underway.

\section{Chiral Transfer from Magnetized Cosmic Rays to Biomolecules}\label{sec-chiralpropa}

The evolution of living organisms is influenced by mutations, which can be caused by cosmic rays which can change the electronic structure of biomolecules, mostly through ionization. Magnetized cosmic rays (i.e. cosmic rays with non-zero lodacity) can affect live and evil molecules slightly differently through the coupling of the lodacity $\cal L$ to the molecular chirality, $\cal M$, which is a pseudoscalar chosen to describe the geometric structure of the molecule, with $|{\cal M}|\le1$ and to have opposite signs for live and evil molecules. 

We confine our attention to very simple, classical and semi-classical models chosen to capture the geometry while ignoring the actual biological and chemical complexity. This is sufficient for our purpose which is to show how, in principle, a chiral bias could be expressed and what factors are essential for it. In addition this approach provides a useful guide for setting up a more realistic calculation based upon quantum chemistry and helps identify special conditions that could lead to a larger bias.

\begin{figure*}
\centering
\includegraphics[scale=0.4]{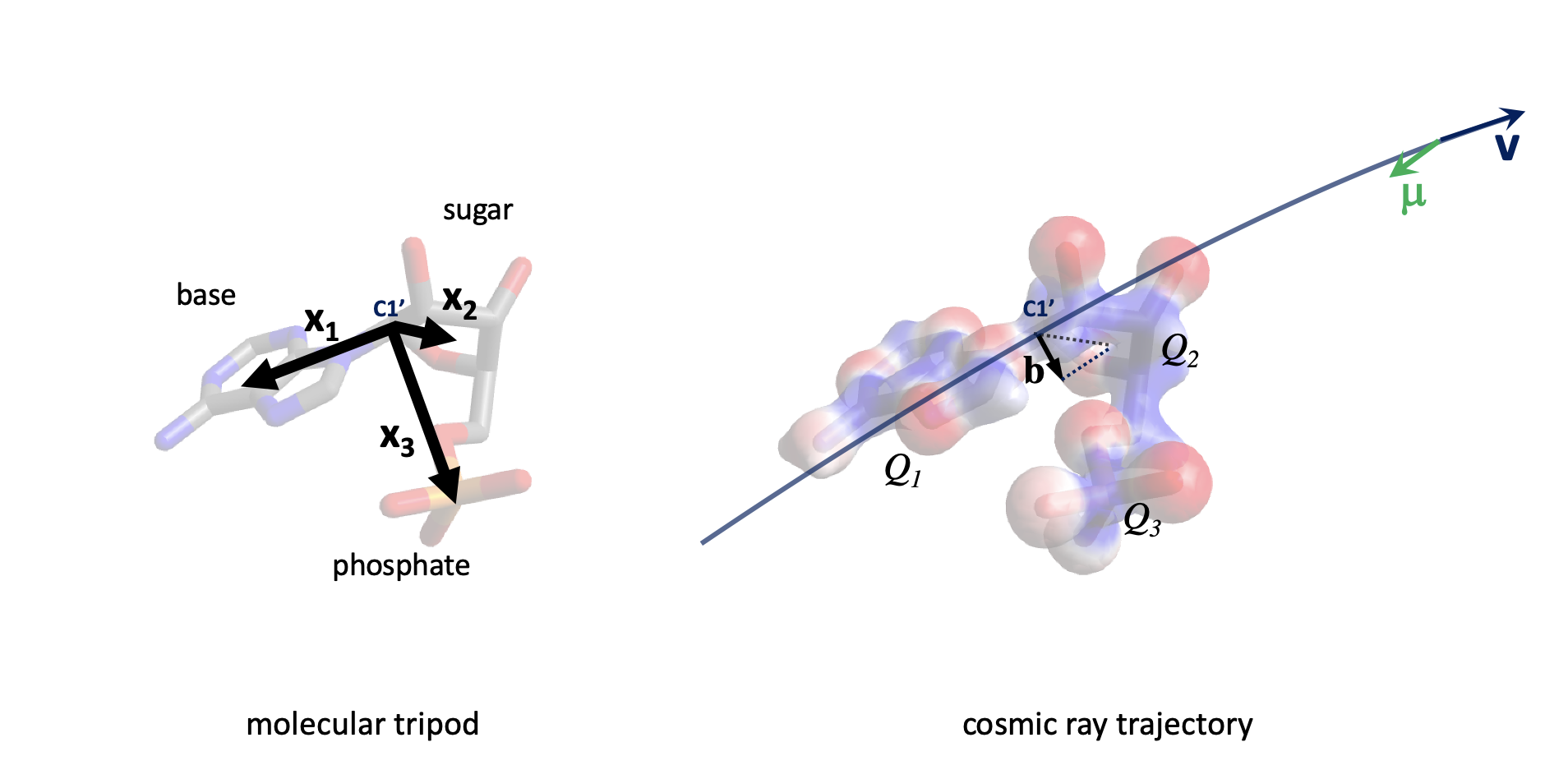}
\caption{\footnotesize Example of electric chirality (tripod model). The electric charge distribution between the three components of nucleic acids (base, sugar, phosphate) of a nucleotide is chiral and the sign of the electric chirality is given by ${\cal M_{\rm tripod}}=\hat{\bf x}_1\times\hat{\bf x}_2\cdot\hat{\bf x}_3$. The electron density (colored by electrostatic potential) is shown on the right; for our simple model we consider the charge distribution of the three groups (base, sugar, phosphate) separately  and study their combined action to the ionization probability of the electrons located around the chiral carbon $C1'$ (which we consider to be the target in our simple model, as it is located between the base - here adenine - and the sugar-phosphate backbone). }
\label{monomer_electric}
\end{figure*}

\subsection{Chiral  monomer}
\subsubsection{Electric chirality (tripod model)}
We first consider a simple model of a small chiral molecule. We suppose that there is a single target site, $O$,  where there is an energy-dependent cross section for inducing a mutation. We then suppose that $O$ is bound to three, non-coplanar, charged sites, representing three atoms or groups at different distances from O. This is the minimum necessary to exhibit chirality. Figure \ref{monomer_electric} shows that this chiral unit can be a simple model of a nucleobase. In this example, $O$ is identified with the atom C1' which bonds the backbone to the base. The chiral bias will then be given by the relative difference in the mutation rate from that for an evil molecule due to the combined action of the three neighboring non-coplanar with $0$ charges (labeled $Q_1$, $Q_2$, $Q_3$ on the figure). 

\paragraph{Electromagnetic force}
For a non-relativistic cosmic ray of charge $q$, mass $M$, magnetic moment $\boldsymbol{\muup}$ and velocity $\bf v$, moving through an electric field $\bf E$, and magnetic field $\bf B$, the electromagnetic force can be written as
\begin{equation}
\textbf F=q(\textbf E+\textbf v\times\textbf B)+\nabla\left[\boldsymbol{\muup}\cdot\left(\textbf B-\frac{\textbf v}{c^2}\times\textbf E\right)\right].
\label{eq:EOM}
\end{equation}

This force is responsible for three types of perturbations which we consider in turn.

\paragraph{Electric deflection}
Let one of these atoms have charge $Q$ and be located at $\textbf x$, relative to $O$. The impact parameter of the passing cosmic ray is ${\bf b}={\bf x}\cdot\hat{\bf v}\hat{\bf v}-{\bf x}$. We measure distance along the trajectory in the direction of motion from closest approach by $z$.
The first order, perpendicular velocity perturbation due to the Coulomb electric field is then given by 
\begin{equation}
\delta {\bf v_\perp}(z)=\frac{qQ{\bf b}}{4\pi\epsilon_0 M  v}\int_{-\infty}^z\frac{dz'}{(b^2+z'^2)^{3/2}}.
\end{equation}
Henceforth, we will measure all lengths in terms of the Bohr radius $a_0=4\pi\epsilon_0\hbar^2/m_e e^2$, all masses in units of $m_e$, all speeds in units of $c$, accelerations in $c^2/a_0$, and all charges in units of $e$. Evaluating the integral, we obtain
\begin{equation}
\delta {\bf v_\perp}(z)=\frac{\alpha^2 q Q}{M b^2 v}\left( 1+\frac z{(b^2+z^2)^{1/2}}\right){\bf b},
\end{equation}
where $\alpha=e^2/4\pi\epsilon_0\hbar c\approx0.0073$ is the fine structure constant.

\paragraph{Magnetic displacement}
In addition, to the electric field, there will be a magnetic field ${\bf B'= -v\times E}$ in the frame of the cosmic-ray and this can interact with its intrinsic magnetic moment $\boldsymbol{\muup}=q e\hbar\hat{\boldsymbol{\muup}}/(2 M m_e)$. This magnetic force, reminiscent of spin-orbit coupling, is given by $\nabla(\boldsymbol{\muup}\cdot \textbf B')=\textbf v \times \boldsymbol{\muup}\cdot \nabla \textbf E$. (The left hand side of this equation is familiar from a Hamiltonian formalism and the right hand side results from recognizing that the field is solenoidal with significant local current.)  Evaluating the electric gradient along the trajectory, the transverse acceleration is
\begin{equation}\label{eq:aperp}
\delta {\bf a_\perp}(z)=\frac{\alpha^3 q Q}{2 (b^2+z^2)^{3/2}}\left({\bf v}(z) \times \hat{\boldsymbol{\muup}}-3\frac{{\bf v}(z) \times \hat{\boldsymbol{\muup}}\cdot {\bf b}}{b^2+z^2} {\bf b} \right),
\end{equation}
where ${\bf v}(z)$ includes the unperturbed and the perturbed velocity.

The associated transverse displacement at $O$ is given by
\begin{equation}\label{eq:deltar}
\delta {\bf r}_\perp^O=-\frac{1}{v^2}\int_{-\infty}^{-\textbf x \cdot \hat{\textbf v}}dz (\textbf x \cdot \hat{\textbf v}+z)\delta {\bf a_\perp}(z),
\end{equation}

\paragraph{Energy shift}
If the cosmic ray undergoes a transverse displacement at $O$, it will acquire a perturbation to its kinetic energy due to the electric potential from an atomic site
\begin{equation}
\delta \ln T= -\frac{2\alpha^2Q}{v^2}{\bf x} \cdot \delta {\bf r}_\perp^O,
\end{equation}

\paragraph{Chiral bias}
Now combine the perturbations. We first suppose that the positrons have a fixed lodacity ${\cal L}=\overline{\hat{\boldsymbol{\muup}}\cdot\hat{\textbf v}}$ and that there is no average perpendicular magnetization, as discussed in Appendix A. This means that the average magnetic moment is along the initial direction of motion. We must apply an electric deflection for there to be an average coupling to the cosmic ray dipole moment. However, this is insufficient for a chiral difference. If we substitute $\delta {\bf v_\perp}(z)$ associated with one atomic site in the expression for $\delta {\bf r}_\perp^O$ (Eq.~(\ref{eq:deltar}) from another site, we obtain a second order displacement at $O$ which is a function of the impact parameter, ${\bf b}$, at the first atomic site at ${\bf x_1}$ and, implicitly, the impact parameter at the second site at ${\bf x_2}$ and also of the initial velocity direction ${\bf \hat{v}}$. The cosmic rays are focused at $O$ and there is a fractional difference in the mutation rate between live and evil molecules given by the relative change in the cosmic ray flux $\delta \ln P=-\partial_{\bf b}\cdot \delta {\bf r}_\perp^O \propto {\cal L}\alpha^5 M^{-3}v^{-3} <\hat{\textbf v} \cdot \hat{\textbf x}_1 \times \hat{\textbf x}_2>$. We call $\delta\ln P$, the chiral bias. If we substitute the evil molecule, the effect has the opposite sign and this, clearly, survives averaging over ${\bf b}$. (When evaluated strictly classically with point charges, $\delta \ln P$ is logarithmically divergent. The divergence is removed if we associate finite de Broglie wavelengths with the particles.) 

If we now average over ${\bf \hat{v}}$ - it suffices to change its sign - this chiral bias vanishes. This is entirely consistent with our expectation. If we only consider second order perturbations, involving two atomic sites in addition to $O$, the interaction cannot be geometrically chiral. So, if the fount is completely isotropic, we need to consider third order perturbations, involving three distinct atomic sites, in addition to $O$, to have the possibility of a chiral coupling. Furthermore, it is apparent that the strength of the perturbations associated with each site is proportional to the scalar charges $Q$, and a third order perturbation to the mutation rate will be simply proportional to the product of these charges, which is unchanged on inversion. It is only their relative locations that matter.

There are several types of third order perturbation. For example, we can use the first order displacement due to the first charge to evaluate the electric field due to the second charge along the perturbed trajectory and compute a second order velocity perturbation to calculate the third order magnetic displacement at $O$.  Alternatively, we can take the second order displacement at $O$ and combine this with the electric field due to a third charge to calculate a third order change in the kinetic energy of the cosmic ray at $O$. We must then sum over all permutations of charge. If the mutation cross section is energy-dependent then there will be an additional chiral bias. All of these terms have the same order, $\delta \ln P\sim\alpha^7{\cal LM_{\rm tripod}}M^{-4} v^{-5}$, where ${\cal M}_{\rm tripod}=\hat{\bf x}_1\times\hat{\bf x}_2\cdot\hat{\textbf x}_3$. In addition it is found that the coefficient vanishes if two of the bond lengths $\bf x$ are equal and the bonds are no  longer chiral. Of course, any of the perturbing charges could also be a target and there could be many more atomic sites. 
 
The effect that we have estimated is manifestly too small ($\sim 10^{-14}$ for relativistic  electrons and $\sim 10^{-22}$ for relativistic muons) to be of interest in the current context. In addition, it demonstrates that if we were to apply it to a variety of pre-biotic molecules then the signs of the individual biass in a set of chemically compatible enantiomers are not guaranteed to be the same and the net chiral bias could be reduced even further.  However, the tripod model is valuable because it directly and explicitly couples the physical chirality of the cosmic ray to the geometrical chirality of the molecule and is highly instructive.

\subsubsection{Electromagnetic chirality}\label{sec-chiralpropa2}
In this second, idealized model, we suppose that, instead of concentrating the charge at three or more points, we distribute electrons smoothly on the surface of a sphere of radius $R$. There may also be a central charge at the origin. The model is semi-classical in the sense that the distribution can be considered as a representation of the expectation of the quantum mechanical charge density. The restriction to the surface of a sphere is not essential --- we could integrate over $R$ --- but it simplifies calculating an actual, chiral bias and suffices to demonstrate electromagnetic chirality and to make some more key points. 

\begin{figure*}
\centering
\includegraphics[scale=0.3]{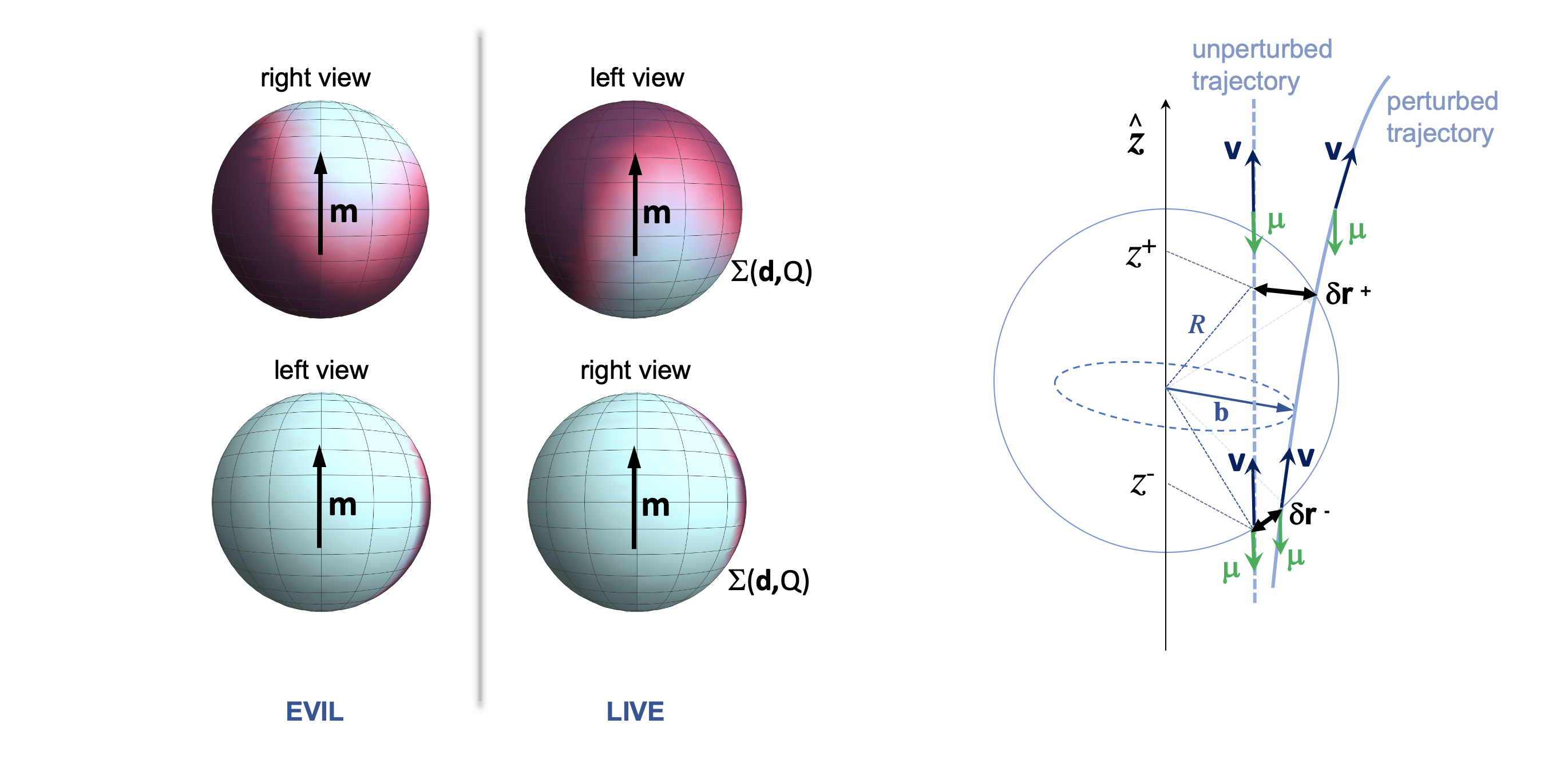}
\caption{\footnotesize Example of electromagnetic chirality. Left: Example of an electric charge distribution  (projected on a sphere) of two biomolecules of opposite chirality, as seen from left and from  right. This simply combines an electric dipole and an electric quadrupole. We need to reflect and rotate by 180 degrees the live molecule to obtain the evil one. The model we discuss in Section~\ref{sec-chiralpropa2} is even simpler, combining electric and magnetic dipoles,  ${\cal M}=\hat{\bf d}.\hat{\bf m}$. The magnetic dipole moment {\bf m} is invariant under parity transformation. Right: Unpertubed vs. pertubed magnetically polarized cosmic ray trajectories through a chiral unit. The unpertubed trajectory is along $z$. The perturbed trajectory due to the chiral molecular field $\textbf B$ is shown. The perturbed cosmic-ray therefore experience a slightly different charge distribution which would lead to a difference in the ionization rate between the two enantiomers.}
\label{monomer_em}
\end{figure*}

We expand the surface charge density for an (arbitrarily assigned) live molecule in terms of multipole moments, chosen to be chiral in combination (Fig.\ref{monomer_em}) and calculate the electric field inside and outside the sphere. We then consider a cosmic ray with velocity $\bf v$ and impact parameter with respect to the origin of the sphere $\bf b$. The linear, transverse displacement at ingress and egress can be expressed as a sum over electric multipoles. We next suppose that the ionization/mutation probability for a cosmic ray traversing the sphere is proportional to the electron densities on the sphere at the actual points of ingress and egress, (designated $-$ and $+$, respectively), correcting for the density gradient. We average over $\bf b$ and $\bf v$ assuming that the cosmic ray flux is uniform and the molecules are isotropically oriented. Next, we repeat the exercise for the evil counterpart molecule and calculated the chiral difference. 

The net result is that there is no purely electrostatic chiral difference for an arbitrary sum of multipoles. This can be understood on quite general grounds because there is no way to combine electric multipole moments electrostatically to form a pseudoscalar. As we demonstrated with the unequal tripod, it is necessary to employ the magnetic field that is created following a frame transformation and this involves the Levi-Civita tensor and, consequently, chirality. Put another way, just because a molecule is geometrically chiral does not ensure a chiral bias; it is necessary to invoke an interaction that couples the molecular chirality to the physical  chirality of the cosmic ray.   

To this end, we now add magnetic multipole moments. In the spirit of our semi-classical approach, we associate these with surface current flowing in the  sphere due to electron orbital angular momentum. Of course, atomic and molecular magnetism is associated more with electron spin but this is unimportant here. It is easy to see that the simplest and strongest molecular chirality combines the electric, $\bf d$, and magnetic, $\bf m$, dipole moments. We call this electromagnetic chirality and it has the simplest definition  ${\cal M}_{\rm em}=<\hat{\bf d}\cdot\hat{\bf m}>$.  We therefore just concentrate on these two multipoles.

\paragraph{Charge density and magnetic field}
The time-averaged surface charge density, $\Sigma$, on the sphere surrounding a point charge $Ze$ at the origin, $O$, is
\begin{equation}\label{eq:surfcharge}
\Sigma=\frac1{4\pi R^2}\left(-Ze+\frac{3{\bf d}\cdot{\bf r}}{R^2}+\dots\right),\quad{\rm for}\,|\textbf r|=R.
\end{equation}
The magnetic field $\bf B$ due to the motion of the electrons on the sphere is 
\begin{equation}
\textbf B=\frac{\muup_0\textbf m}{2\pi R^3}+\dots,\;r<R,\quad=\frac{\muup_0}{4\pi}\left(\frac{3(\textbf m\cdot\textbf r)\textbf r}{r^5}-\frac{\textbf m}{r^3}+\dots\right),\;r>R.
\end{equation}

\paragraph{ Cosmic Ray Path}
We now consider the path of a single cosmic ray with impact parameter \textbf b with respect to the center $O$ of a live molecule, as seen in Fig.\ref{monomer_em}. The (classical) force acting on the cosmic ray is given by Eq.~\ref{eq:EOM}
and we need the transverse displacement as the cosmic ray enters and leaves the sphere. We introduce the coordinate $z=\textbf r\cdot\hat{\textbf v}$, so that ingress and egress are at ${\textbf r}^\mp=\textbf b + z^\mp\hat{\textbf v}$ with $z^\mp=\mp(R^2-b^2)^{1/2}$. We continue to assume that $\boldsymbol{\muup}$ is on average antiparallel to $\hat{\bf v}$. The largest chiral interaction is with $\bf B$ and there is a linear perturbative force of $\mu_z\nabla B_z$. Prior to ingress, we use $\nabla\times{\bf B}=0$  to find that the velocity perturbation is $\delta{\bf v}=\muup_z{\bf B}/Mv$.

We can now calculate the displacement perpendicular to the unperturbed path at ingress. (This is not same as the displacement at a fixed time.) This is given to first order by
\begin{align}
\delta\textbf r^-_\perp=&
\frac{\muup_{\rm z}}{2T}\int_{-\infty}^{z^-}dz\left(\textbf B-(\hat{\textbf v}\cdot\textbf B)\hat{\textbf v}\right),\nonumber\\
=&\frac{\muup_0\muup_{\rm z}}{8\pi TR^2}
\left[\left(\frac{1-(1+\eta)(1-\eta)^{1/2}}\eta\right)(\textbf m\cdot\hat{\textbf b})\hat{\textbf b}-\eta^{1/2}(\textbf m\cdot\hat{\textbf v})\hat{\textbf b}\right.\nonumber\\
&\left.-\left(\frac{1-(1-\eta)^{1/2}}\eta\right)(\textbf m\cdot\hat{\textbf v}\times\hat{\textbf b})\hat{\textbf v}\times\hat{\textbf b}\right].
\end{align}
where $T=Mv^2/2$ and $\eta=b^2/R^2$.

The velocity perturbation immediately after ingress has to take account of the impulse due to the current flowing on the surface of the sphere. However, there is no additional force as the interior magnetic field is uniform. The chiral part of the transverse displacement at egress can then  be shown to be 

\begin{align}
\delta\textbf r^+_\perp = &\delta\textbf r_\perp^-+\frac{(z_+-z_-)\delta\textbf v_\perp^+}v\nonumber\\
=&\frac{\muup_0\muup_{\rm z}}{8\pi T R^2}
\left[\left(\frac{1-(1+3\eta)(1-\eta)^{1/2}}\eta\right)(\textbf m\cdot\hat{\textbf b})\hat{\textbf b}-7\eta^{1/2}(\textbf m\cdot\hat{\textbf v})\hat{\textbf b}\right.\nonumber\\
&\left.-\left(\frac{1-(1+4\eta)(1-\eta)^{1/2}+6\eta(1-\eta)}\eta\right)(\textbf m\cdot\hat{\textbf v}\times\hat{\textbf b})\hat{\textbf v}\times\hat{\textbf b}\right].
\end{align}

\paragraph{Ionization Rate}
The classical ionization cross section is 
\begin{equation}
\sigma_{\rm ion}=4Z\pi a_0^2\left(\frac{I_H}I\right)\left(\frac{I_H}T\right)\sim3.5\times10^{-21}T_{\rm keV}^{-1}\,{\rm m}^2\,.
\label{Ionization Cross_Section}
\end{equation} 
The probability that a cosmic ray incident upon an atom, idealized as a charged sphere, will create an ionization is therefore $P_{\rm ion}\sim\sigma_{\rm ion}/\pi R^2\sim0.2\,T_{\rm keV}^{-1}$. Direct measurements below $\sim1\,{\rm keV}$ give cross sections lower by factors up to ten and a slower decline with increasing kinetic energy \citep{kim2002}. This reflects the fact that more tightly bound electrons can be ionized as $T$ increases as well as quantum mechanical effects. Again, this is unimportant for our limited purpose. In addition to its transverse displacement, a cosmic ray will have a slightly different energy and cross section as it crosses the sphere and there is an associated chiral bias. This turns out to be subdominant in our model and we ignore it although it is likely to be significant in a more realistic description. 

\paragraph{ Chiral bias}
We have computed the first order deflection at ingress and egress. By itself, this leads to no net change in the ionization rate. However, the deflection results in the cosmic ray encountering a slightly different surface density of electrons due to the gradient in the electron density within the sphere. The second order change in the relative ionization rate, the chiral bias, is then given by
\begin{equation}
\delta\ln P= - (\delta\textbf r_\perp^-\cdot\nabla_\perp\ln\Sigma^- +\delta\textbf r_\perp^+\cdot\nabla_\perp\ln\Sigma^+),
\end{equation}
where the perpendicular, logarithmic gradient in the relative surface charge density at ingress and at egress is
\begin{align}
\nabla_\perp\ln\Sigma^\mp=
-\frac3{ZR^2}\left((1-\eta)^{1/2}\eta(\textbf d\cdot\hat{\textbf b})\hat{\textbf b}\pm\eta^{1/2}(\textbf d\cdot\hat{\textbf v})\hat{\textbf b}
+\textbf d\cdot\hat{\textbf v}\times\hat{\textbf b}\,\hat{\textbf v}\times\hat{\textbf b}\right),
\end{align}
and we have used Eq.~(\ref{eq:surfcharge}). (There is no gradient in the monopolar surface density and any quadrupolar term does not survive averaging.)

So far, we have considered one molecule and one cosmic ray. We must now average over direction. The simplest assumption to make is that the cosmic ray flux is isotropic with respect to the molecule. This still allows the cosmic rays to be anisotropic if, as is likely, the molecules are randomly oriented, for example in water. (We note that there are circumstances when orientation biases may be present and these could lead to a larger chiral bias.)  In order to carry out the angle average, we first note that any term contributing to $\delta\ln P$ that is odd in  $\hat{\textbf v}$ can be dropped as it will be canceled by the effect of a cosmic ray with the opposite velocity or impact parameter. We then average the remaining terms over azimuth, perpendicular to \textbf v using $<\textbf m\cdot\hat{\textbf b}\;\textbf d\cdot\hat{\textbf b}>=\textbf m\cdot\textbf d/2$ etc. Finally, we average $\eta$ over a unit circle.  The final step is to average $\hat{\textbf v}$ over the surface of a unit sphere. Averages of the form $<\textbf u\cdot\hat{\textbf v}\,\textbf w\cdot\hat{\textbf v}>$ become $\textbf u\cdot\textbf w/3$. After performing these integrals and averages, we obtain a chiral bias
\begin{equation}\label{eq:probion}
\delta\ln P=-0.98\alpha^4Z^{-1}dm{\cal LM}M^{-2}v^{-2}
\end{equation}
where the dipole moment of the chiral unit, $d$, is measured in units of D $\equiv0.37 ea_0$ and the magnetic moment, $m$, is in $\mu_{\rm B}$.

\paragraph{Magnetic moment}
Electromagnetic chirality can lead to a relatively large chiral bias \citep[cf.][]{zeldovich1977}. However it requires the magnetic moment in a small chiral unit to be aligned systematically with the dipole moment. The permanent magnetic moments of small biological molecules have been studied less than their electric dipole moments. They should exist when there are unpaired electron spins and they contribute to the paramagnetic susceptibility. However, there does not seem to be a good atomic physics reason why they should align with the electric dipole moment. Another way of expressing this is to say that the volume integral of the relativistic invariant ${\bf E}\cdot{\bf B}$ over all space is $-Z_0^2{\bf d}\cdot{\bf m}/3\pi R^3$, where $Z_0$ is the impedance of free space and a reason has to be found why this should be non-zero.

\subsection{Helical biopolymer}
We now turn to an idealization of a biopolymer. We hypothesize that single-stranded, naturally twisted  biopolymers, that had some limited capacity to replicate, albeit with a high frequency of errors/mutations, were the prime genetic agent when the transition to life occurred. We argue that, if this were the case, these long, helical molecules represent a more likely candidate for an explanation of homochirality than the much smaller molecules that were present during a pre-biotic epoch.  These simpler and more primitive helical precursors contrast with the highly-evolved  DNA and RNA of today, which replicate relatively efficiently and with much greater fidelity.  An early, twisted biopolymer should exhibit no chemical preference for handedness. However, the twist is likely to exhibit long range order along the polymer \citep{keating2011}. A single cosmic ray will only interact with a short segment of the molecule and it is the generic, geometrical disposition of the electrical charge and field within this segment that confers the sign of the chiral preference. This should be common to most helical biopolymers and leads to a much larger chiral bias than the tripod model.

\subsubsection{Electric chirality (barber pole model)}
A RNA-like molecule comprises a chain of nucleotides that spirals like a barber pole and can be modeled by an electrostatic potential, $\Phi=R_0(r)+R_1(r)\cos[kz-m(\varphi-\varphi_0)]$, with radial and helical terms \citep[cf.][]{wagner1997}. As the helix is assumed to be single-stranded, we set the azimuthal quantum number to $m=1$ for a live progenitor and to $m=-1$ for the evil counterpart. (For a double stranded molecule like DNA, $|m|=2$ is a better approximation but there will be similar effects.) As lengths are measured in units of the Bohr radius, $k\sim0.1$ for modern RNA and $\Phi$ is measured in units of $e/(4\pi\epsilon_0 a_0)$.) One way to show formally that this model is chiral is to suppose that $R_1(r)$ has a maximum at $r_{\rm max}$. Construct a radius vector ${\bf r}_0$ from the axis to this maximum at $z=0$.  Displace the origin of this vector a distance $\bf z$  along the axis until the radius vector, now ${\bf r}_1$, has turned though $\pm\pi/2$. 
The quantity ${\bf \hat{r}_0 \times \hat{r}_1 \cdot \hat{z}}$ involves the Levi-Civita tensor, and is therefore chiral. It is unchanged under the transformation ${\bf \hat{z}}\rightarrow -{\bf \hat{z}}$. Note that if we regard the two radius vectors as defining a tetrahedron, the equal edges do not share a common vertex. This differentiates the barber pole from the equal tripod which is non-chiral.

\begin{figure*}
\centering
\includegraphics[scale=0.32]{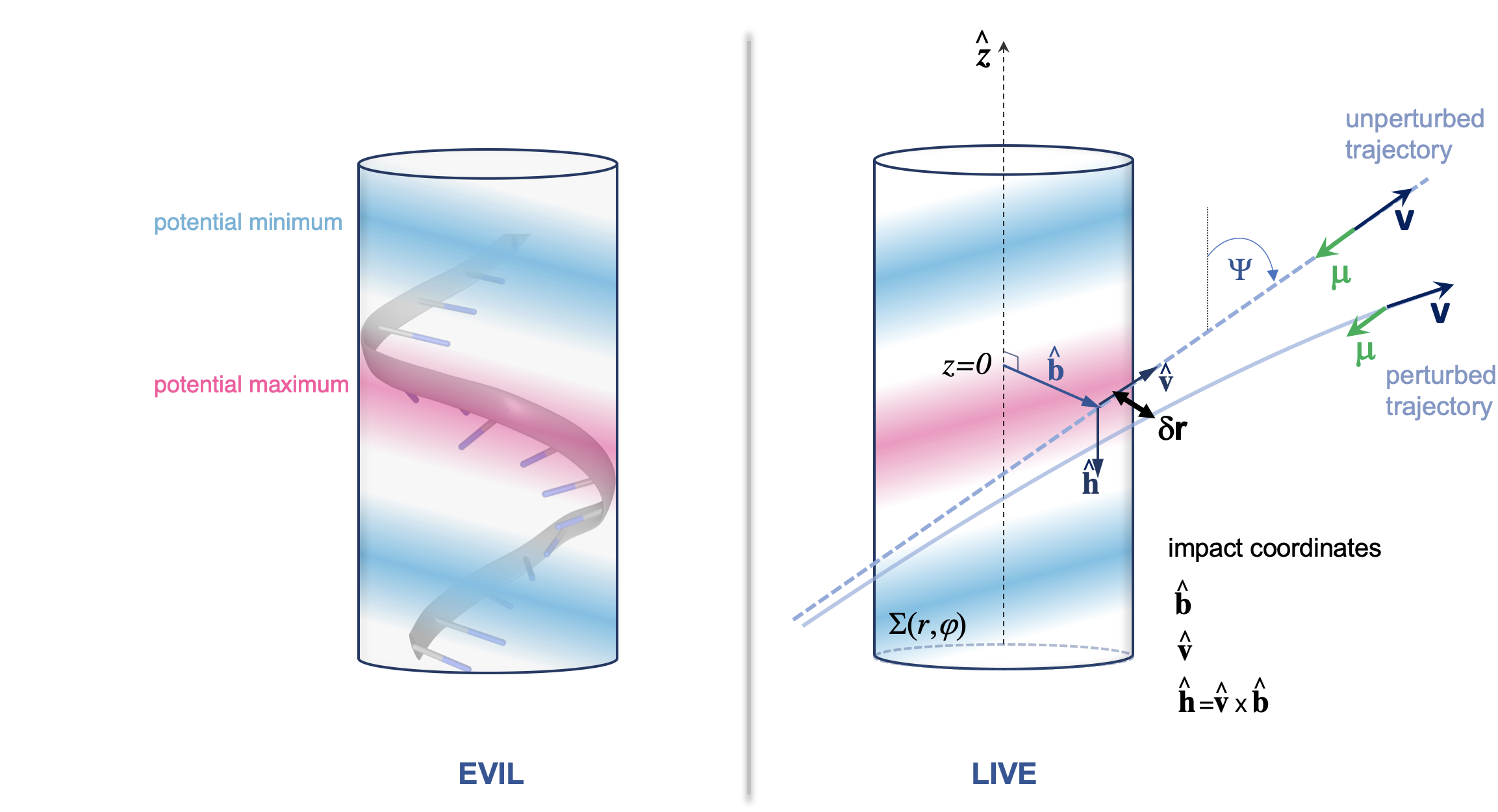}
\caption{\footnotesize Example of electric chirality (barber pole model). The electric charge distribution   of two biopolymers of opposite chirality, ($\Sigma (r,\varphi)$ projected onto a cylinder) is shown, together with the unperturbed vs. perturbed trajectory of a magnetically polarized cosmic ray interacting with the molecule.}
\label{barberpole}
\end{figure*}

\paragraph{Electric deflection}
Consider a cosmic ray with charge $qe$, mass $M m_e$ and velocity ${\bf {v}}$ (measured in units of c) making an acute angle $\Psi$ with ${\bf \hat{z}}$, as shown in Fig.~\ref{barberpole}. Let the cosmic ray have impact parameter ${\bf {b}}$ lying in the $z=0$ plane and set $\varphi=0$. Its unperturbed trajectory is ${\bf s}={\bf b}+z\sec\Psi{\bf \hat{v}}$. 
 
The perpendicular velocity perturbation due to the Coulomb electric field is given by

\begin{equation}
\delta {\bf v_\perp}=-\frac{q \alpha^2}{ M  v\cos\Psi}\int_{-\infty}^{z}dz'(\nabla\Phi)_\perp.
\end{equation}

\paragraph{Magnetic displacement}
Just as with the tripod model, we can calculate the second order perturbation to the transverse magnetic acceleration in direct analogy to Eq.~(\ref{eq:aperp})
The magnetic force is still given by $\nabla(\boldsymbol{\muup}\cdot \textbf B')=\textbf v \times \boldsymbol{\muup}\cdot \nabla \textbf E$. Evaluating the electric gradient along the trajectory, the transverse acceleration is

\begin{equation}
\delta {\bf a_\perp}=\left(\frac{q\alpha^3{\hat{\boldsymbol{\muup}} \times{\bf v}(z)}}{2M^2}\cdot \nabla\right)\nabla\Phi\,.
\end{equation}

Hence, using our definition of lodacity,
\begin{equation}
\overline{\delta {\bf r_\perp}}=\frac{1}{(v\cos\Psi)^2}\int_{-\infty}^{z}dz'(z-z')\overline{\delta {\bf a_\perp}}=-\frac{q\alpha^5{\cal L}}{2(M v \cos\Psi)^3}\int_{-\infty}^{z}dz'\int_{-\infty}^{z'}dz''(z-z'') ({\bf \hat{v}}\cdot \nabla\Phi(z')\times \nabla)\nabla\Phi(z'')
\end{equation}
where the bar denotes an average over the cosmic ray magnetic moment.

\paragraph{Chiral bias}
We now make a change from the assumption that we made when we considered electromagnetic chirality. Instead of assuming that the density of ionizable electrons is essentially that of the perturbing electrical charge, for a helix, we suppose, instead, that the effective density is distinct, while sharing the same symmetry as the electrostatic potential. The reason for doing this in the context of RNA and its possible progenitors is that we do not understand the path from ionization to mutation. It may be more important to break bonds in the central bases. Alternatively, the outer sugar backbone may be more relevant.

In order to include this freedom,  we introduce a "mutability", $\kappa$, (probability per unit length of cosmic ray trajectory for mutation). The lodal contribution to the probability of mutation is then
\begin{equation}
{\delta P}=\int_{-\infty}^{\infty}dz\, \overline{\delta {\bf r_\perp}}(z)\cdot \nabla \kappa (z)=-\frac{q\alpha^5{\cal L}}{2(M v \cos\Psi)^3}\int_{-\infty}^{\infty}dz\int_{-\infty}^{z}dz'\int_{-\infty}^{z'}dz''(z-z'') {\bf \hat{v}}\cdot \nabla\Phi(z')\times \nabla\nabla\Phi(z'')\cdot \nabla \kappa (z)
\end{equation}
where the only derivatives we need are perpendicular to ${\bf\hat v}$ along ${\bf \hat{b}}$, ${\bf \hat{h}}$. Note that the effect has opposite sign for the two signs of cosmic ray charges and the same negative sense of lodacity.\\

We adopt a general, separable potential of the form $\Phi=R_0(r)+R_1(r)\cos[kz-m(\varphi-\varphi_0)]$, with radial and helical terms.  The associated charge density contributed by the combination of the nuclei, inner shell electrons and binding electrons is given by $\rho=-\epsilon_0\nabla^2\Phi$ and should represent the actual distribution in a realistic model. Likewise, we assume $\kappa=K_0(r)+K_1(r)\cos[kz-m(\varphi-\varphi_0)]$. ${\delta P}$ can be considered as a triple integral over $z$, $z'$, $z''$, proportional to a sum over terms containing factors describing the mutability gradient, the electric field and the electric field gradient contributing to the pseudoscalar ${\cal F}={\bf \hat{v}}\cdot \nabla\Phi(z')\times \nabla\nabla\Phi(z'')\cdot \nabla \kappa (z)$.

Next, expanding all the trigonometric factors in the expression for ${\delta P}_{\rm live}$, ($m=1$). The terms in ${\cal F}$ involve products of the radial functions $R_0(r), R_1(r), K_0(r), K_1(r)$ and derivatives. (There are nearly 50,000 terms!) Likewise, for the evil molecule with $m$=-1. We then  subtract the evil from the live terms. The next step is to average over $\varphi_0$ and then average the velocity with inclination $\Psi$ over the surface of a sphere. This is made difficult because $\Psi$ appears in the arguments of the $K$ and $R$ functions. We must also average the impact parameter, $b$ over the circumference of a circle. 

The integrals over $z$, $z'$, $z''$ require substituting $r=(b^2+z^2 \tan^2 \Psi)^{1/2}$, $\varphi=\tan^{-1}((z \tan\Psi)/b)$, etc. (In principle, derivatives $R_0(r), R_1(r), K_0(r), K_1(r)$ of can be removed through integration by parts.) In order to complete the calculation we would have to perform a five dimensional integration for each specific choice of the $R$ and $K$ functions. An important general conclusion is that, after performing the averages, all of the chiral terms combine three factors with one factor being a radial function, either $R_0(r)$ or $K_0(r)$, and two of them being helical functions chosen from $R_1(r)$, $K_1(r)$. In practice it is better to perform these calculations by imitating the cosmic rays and performing a Monte Carlo sampling of isotropic cosmic ray trajectories. However, the analysis so far suffices to demonstrate that there is a finite chiral bias and what it depends upon. 

It is not obvious which handedness is preferred, but it is probably generic, depending mostly on the gross charge distribution. It appears that the dominant chiral combination of functions is $R_1R_1K_0$ and derivatives. Note that it is not, in practice, necessary for the mutability to be helical. If this is indeed the case, then it appears to be generically true that the live chiral bias follows from the mutability being associated with ionizing the central bases instead of the sugar-phosphate backbone. This is under investigation.

The best estimate for the chiral bias for mutation is $\delta\ln P\sim \alpha^5{\cal LM}M^{-3}v^{-3}$. Several comments can be made. For a given speed and lodacity, muons are $\sim 10^{-7}$  times as effective as positrons or electrons and we therefore emphasize the latter. Clearly lower speed particles are also more effective. However, the lodacity of the initial cosmic rays is ${\cal L}\propto v$ so the chiral bias of the primaries is $\sim \alpha^5 v^{-2}$. This can be as large as $\sim 10^{-7}{\cal L}$.

Secondary electrons will generally be unpolarized (although spin-dependent interference terms in the scattering cross section of identical electrons \citep{messiah1981} might confer some persistence of the lodacity). To address this requires a more careful shower simulation. Also it should be emphasized that a classical approach surely fails when $v$ declines towards $\alpha$, the characteristic speed of a molecular electron. Under these circumstances, a better approach is to solve for the electron orbitals in a simple idealization of the barber pole and to compute the matrix elements and transition probabilities for collisional excitation and ionization. This should exhibit a qualitatively similar chiral bias to the semi-classical calculation we have sketched here.

\subsubsection{Electromagnetic chirality}
We have not considered, yet, the possibility of electromagnetic chirality of helical biopolymers. External electric and magnetic fields would influence the conformational flexibility of nucleic acids and  hence affect the molecular chiral quantity ${\cal M}$. Although it has been shown that RNA and DNA are ferroelectric, nucleic acids do not show strong permanent magnetic moments (at least in their neutral form). Local correlations between neighboring bases could, in principle, cause local $\hat{\bf d}\cdot \hat{\bf m}$ but this  is far from certain. However, when a magnetic field is applied, nucleic acids become magnetic. This is due to the aromatic rings in the bases. In the presence of a magnetic field, magnetic moments perpendicular to the plane of the bases are induced by  the so-called ring currents. These induced magnetic moments would be localized towards the center of the helix, where the bases are located. It would surely induce a preferred orientation for the biopolymers (likely the same effect as the influence of a ferromagnetic fount) and this could change the strength of the molecular chirality, but not its sign which is related to the handedness of the helix. 

We emphasize that it is the geometrical, helical structure of the biopolymers that appears to have the durability and strength to define and sustain  a universal handedness. The electromagnetic chirality $\hat{\bf d}\cdot\hat{\bf m}$ of a monomer can depend on the pH of the solvent and cannot lead to a universal sign. It is of interest to investigate the effect of magnetic moments in an helical configuration (they would also spiral around the helix axis), superimposed with the electric molecular chirality in the barber pole model.

\begin{figure*}
\centering
\includegraphics[scale=0.23]{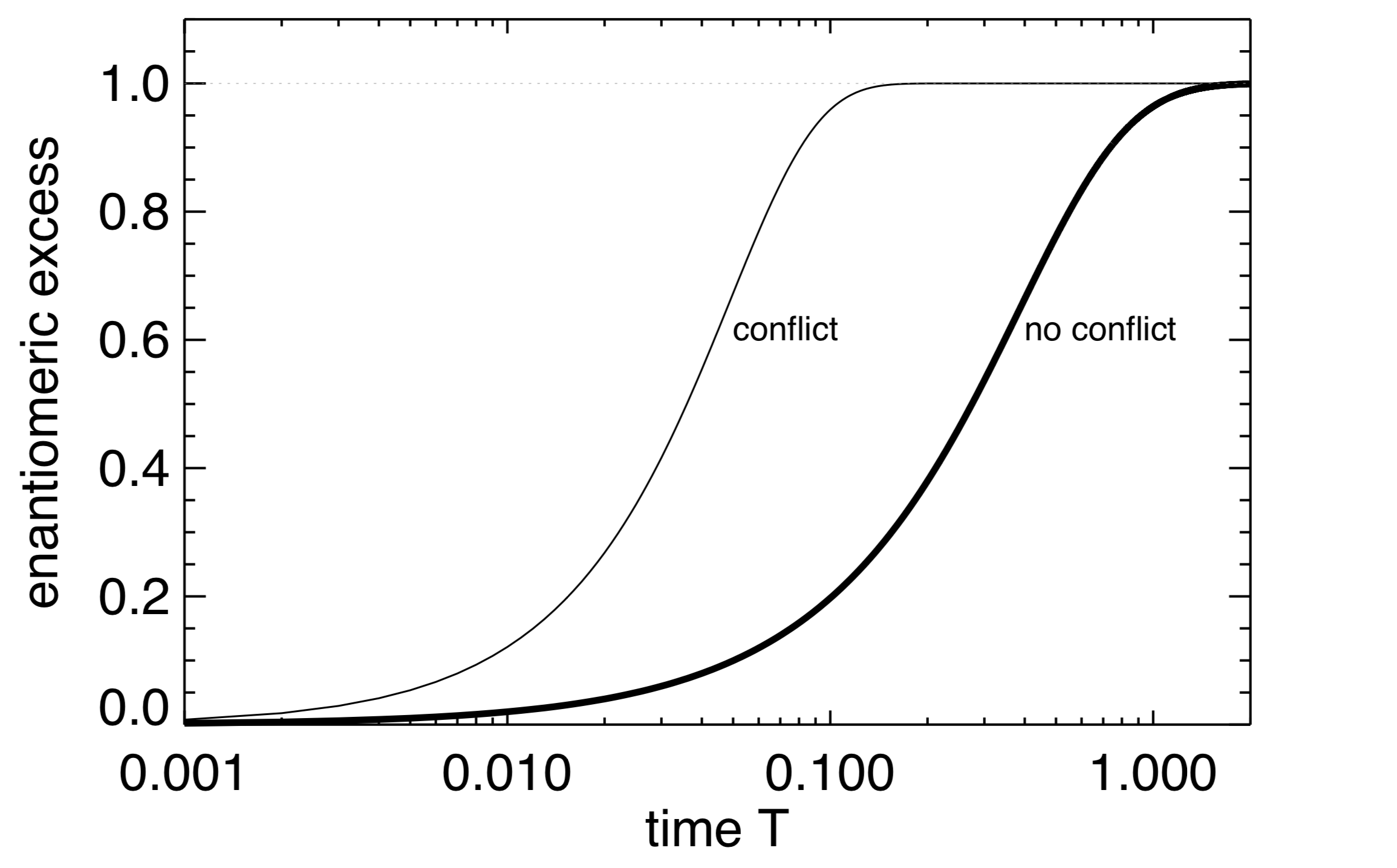}
\caption{\footnotesize Solutions of Eqs.\ref{logistic}-\ref{logistic2} for $x=0$ (thick line) and $x= 10^4$ (thin line). For $x = 10^4$, the homochiralization time scale divided by $\sim10$. We start at $T=0$ with a racemic mixture (as this is not a model of amplification of a small initial fluctuation by a mutual antagonism; in fact, the antagonism is not necessary). The homochiralization is a result of the cosmic rays lodacity that can impose a small, but persistent, chiral bias.}
\label{fig:mirrorbreak}
\end{figure*}

\section{Breaking the biological mirror}\label{sec-evolampli}

We have shown that there are interactions that can couple the lodacity of cosmic rays to the molecular chirality of the first, vital molecules. The biases we have estimated are all very small, $\sim10^{-7}$ for keV electrons (times the lodacity and the fractional difference between positive and negative charges in the barber pole model), although larger biases may be found through including other factors in the interaction. We now turn to considering how this small enantioselectivity might evolve over time to homochirality.  

Enantioselective auto-catalysis, where molecules  of the same chirality  catalyze their own production while inhibiting the formation of their mirror-image \citep{frank1953}, has served as an important model because it exhibit some of the features of life, {\textit i.e.}, self-replication. However, living organisms have the ability not only to self-replicate but to increase their complexity, whenever complexity is beneficial to their survival. This must be the case for any early life form not yet well adapted to its environment.  The  evolutionary change is based on the accumulation of many mutations with small effects.

The way a biopolymer folds into space determines its biological function. Clay minerals, present in the fount, may have catalyzed the polymerization of the first biopolymers; they can also protect the  bases adsorbed on their surface from radiation \citep{biondi2007}. They, too, could play a role in the homochiralisation process, enhancing the chiral selection \citep[see e.g. ][for a review]{hazen2010}. It has also been shown that clay minerals can protect the building blocks of biomolecules against the effects of high energy radiation \citep{guzman2009}. 

We do not deal here with the chemical pathway needed for the assembly of the first polymers \citep{leslie2004}. In the absence of a chiral driving force, a prebiotic chemical reaction necessarily yields a racemic state; prebiotic pathways leading to a non-racemic (but not necessarily homochiral) state have been explored  invoking chiral catalysts, that need to be be present in the fount to induce a bias \citep{soai1995, breslow2010}. Instead we propose that prebiotic chemistry produces both chiral versions of the molecular ingredients of life (\textit{i.e.} helical polymers capable of self-replication), and that at some stage in the earliest development of  biomolecules, a small difference in the mutation rate, attributable to the lodacity of the cosmic rays, gives a chiral bias to the live genetic polymers over their evil counterparts. 
 
Cosmic rays are generally recognized as agents of natural selection. At modest intensity, which interests us here, they promote mutation and natural exploration of biochemical and evolutionary pathways; when the intensity is high, they will be destructive and will create sterile environments.  It seems that cosmic radiation also affects the growth rate of the living organisms; for example, it has been reported that during episodes of high cosmic-ray flux and cold climate there is an enhancement of biological productivity \citep{svensmark2006} although this relationship is controversial. Conversely, radiation deprivation has been reported to inhibit bacterial growth \citep[e.g.][]{castillo2015}.

Growth rates and mutation rates are correlated functions. When the growth rate is low, the probability to accumulate an adaptive mutation is strongly limited. To demonstrate the effect we therefore assume a simple relation between the growth rate and the rate of mutation in the nucleobase sequence, $g(M)={\cal C}\sigma_M F$,  where ${\cal C}$ is a positive constant, $\sigma_M$ is the mutation cross section (which depends on ionization and excitation of the nucleobases) and $F$ the cosmic ray flux. While it is reasonable to postulate that cosmic rays increased the rate of genetic mutations in proto-lifeforms, the exact  relationship between  the radiation dose and the mutation rate is unknown. For sake of simplicity, we assumed a linear dose response. Our argument that a difference in the mutation rate of live and evil organisms would lead to homochirality is not premised on the linear proportionality of $g$ on the cosmic ray flux $F$. However,  the timescale at which homochiralization occurs, does depend on the relationship between $g$ and $F$, so the reader should keep in mind that our simple, working hypothesis might be inadequate to estimate the homochiralization timescale.

When considering living organisms, the definition of "enantiomeric excess" is more subtle, because the living organisms are never the same molecular entities at a given time. However, even if the genetic information evolves, the chirality of the nucleotides (which is related to the handedness of the sugar) is maintained. We denote by $N_{\rm live}$ and $N_{\rm evil}$ the number of live and evil molecules, respectively. The evolution of the two populations is given by:
\begin{eqnarray}
    \frac{d\ln N_{\rm live}}{dT}&=& 1+\frac{\delta M}{2}-x N_{\rm evil}\,,\label{logistic}\\
    \frac{d\ln N_{\rm evil}}{dT}&=&1-\frac{\delta M}{2}-x N_{\rm live}\,.
    \label{logistic2}
\end{eqnarray}
where 
\begin{itemize}
\item[$\bullet$]$T (\sigma_M, {\cal C}, F)$ is the time multiplied by the growth rate;
\item[$\bullet$] $x$ is the  antagonism rate divided by the growth rate;
\item[$\bullet$]$\delta M({\cal L, \cal M})\sim \delta\ln P$ is the relative   difference in the mutation rates estimated in the previous section. 
 \end{itemize}

With zero conflict, {\it i.e.} $x=0$, the enantiomeric excess is $e.e.=(N_{\rm live}-N_{\rm evil})/(N_{\rm live}+N_{\rm evil})=1-2[1+ N_{\rm live,0}/N_{\rm evil,0}\exp(\delta M T)]^{-1}$. If we start with a racemic mixture (which is likely to be the case as shown by laboratory experiments) then ${N_{\rm live,0}}/{N_{\rm evil,0}}=1$ and  $e.e.={\rm tanh}(\delta M T/2)$.  Homochiralisation  occurs when $T \sim4\delta M^{-1}$.  The antagonism is not necessary. It shortens the time scale at which homochiralisation occurs, as can  be seen in Fig.\ref{fig:mirrorbreak}. 

\end{document}